\documentclass[preprint,preprintnumbers,amsmath,amssymb,citeautoscript]{revtex4}
\usepackage{graphicx}% Include figure files
\usepackage{dcolumn}%Align table columns on decimal point
\usepackage{bm}% bold math
\usepackage{color}
\setcitestyle{super}

\begin{document}

%\preprint APS/123-QED}

\title{{\it Ab Initio} Potential Energy Surfaces and \\
Quantum Dynamics of Rotational Inelastic Processes in the \\ 
H$^+$ Collision with CS ($^1\Sigma^+$)}
\author{Rajwant Kaur}
\author{T. J. Dhilip Kumar}
\email{dhilip@iitrpr.ac.in}   
\affiliation{Department of Chemistry\\
Indian Institute of Technology Ropar\\
Rupnagar 140001, India\\
}
%{\footnote{Present address}\\

\date{\today}% It is always \today, today,

\begin{abstract}
Rate coefficient for state-to-state rotational transitions in 
H$^+$ collision with CS has been obtained using accurate quantum dynamical close-coupling 
calculations to interpret microwave astronomical observations. 
Accurate three dimensional {\it ab initio} potential energy surfaces have been computed 
for the ground state and low-lying excited states 
of H$^+$ $-$ CS  system using internally contracted MRCI method and
aug-cc-pVQZ basis sets. 
Rotational excitation and deexcitation integral cross-sections are computed at low 
and ultra low collision energies, respectively. 
Resonances have been observed at very low energies typically below 50 cm$^{-1}$. 
Among all the transitions, $\Delta$$j$=$+$1 and $\Delta$$j$=$-$1 are found to be predominant for excitation and deexcitation, respectively. 
Deexcitation cross-section in the ultracold region is found to obey Wigner's threshold law. % where only the $s$-wave scattering contribute 
%and the cross-section vary inversely with the relative velocity. 
The magnitude of state-to-state excitation rate obtained is maximum for $j'$=1 in the temperature range 2$-$240 K while minimum for deexcitation in 
ultracold region. 
The rotational excitation cross-section obtained using vibrationally averaged potential show rotational rainbow maximum for $j'$=2 state.
From simple unimolecular kinetics, the mean lifetime of rotationally excited CS trap is estimated to be 550 ns
due to the H$^+$ collision at microkelvin temperature enabling precise spectroscopic measurement and studying molecular properties
near quantum degeneracy.

\end{abstract}

%\pacs{}% PACS, the Physics and astronomy
%\keywords{} 
%Use showkeys class option if keyword
\maketitle

\section{\label{sec:intro}INTRODUCTION}
Chemical processes in the interstellar medium occurring
due to the ion-neutral collisions play a fundamental
role in the areas of molecular physics and astrophysics.
Proton collision with diatomic molecule is important in atmospheric
chemistry and combustion processes. 
Various bound protonated molecular ions, such as H$_3^+$,
HCO$^+$, HCS$^+$, etc., have been identified in the interstellar
medium through radio-astronomical spectra.\cite{herbst,weaver} 
Inelastic and charge transfer processes are responsible for the formation
of these protonated ionic species.
Many sulfur containing diatomic and polyatomic molecules have been detected %till date 
in dense interstellar clouds.
Among them, carbon monosulfide (CS) discovered in 1971 in the interstellar clouds % \cite{penzias} 
and thioformyl cation, HCS$^+$, detected in 1981 
have attracted attention in understanding
the sulfur chemistry of the interstellar clouds.\cite{penzias,thaddeus} 
CS is the first sulfur containing molecule detected in the interstellar space. It is a tracer
gas in several regions of interstellar medium in our galaxy as well as in external galaxies.
%CS molecule is an important and major source for producing HCS$^+$. 
%Abundance of HCS$^+$ depends on CS$^+$ which in turn depends upon CS abundance.\cite{mcallister} 
Collision of CS molecule with
neutral species have been studied with most abundant molecules, He and H$_2$.\cite{lique2,alpizar}
%HCS$^+$, is the precursor of CS and its formation is due to 
%high proton affinity of CS.
HCS$^+$ detection in dense interstellar clouds is due to large CS proton affinity and  
CS on reaction with protonated H$_3^+$ species favor HCS$^+$ formation.\cite{mcallister,clary}\\

Several chemical schemes have been proposed for the formation and dissociation of HCS$^+$ with its dissociative
recombination mechanism and the results are validated by rate-coefficient calculations. 
Branching ratios and absolute
cross-sections have been measured for the dissociative recombination of HCS$^+$.\cite{montaigne}
Gerones {\it et al.} have analyzed the HCS$^+$ formation and its high stability over the range of 
photon energies.\cite{gerones1,gerones2}  
In the past {\it ab initio} studies have been performed to obtain the spectroscopic properties of HCS$^+$ with its optimized structural 
parameters.\cite{botschwina}
Configuration interaction (CI) method with triple-zeta basis sets has been employed to study  HCS$^+$/HSC$^+$
isomerization in 1985.\cite{pope}
In 1991, Talaty {\it et al.} reported that the HCS$^+$ isomer is linear and global minimum while its HSC$^+$ isomer is bent and higher in energy.\cite{talaty}
Molecular properties, isomerization, and energetics
of neutral CS, HCS/HSC and HCS$^+$/HSC$^+$ systems have been studied by Puzzarini.\cite{puzzarini}
Cotton {\it et al.} have performed structural and spectroscopic studies of CS-HCS$^+$ van der Waals complex.\cite{cotton}
The structural parameters, harmonic vibrational frequencies and charge distribution data of noble gas (Ng) inserted in the HCS$^+$ ion resulting in 
HNgCS$^+$ has been studied at MP2, DFT and CCSD(T) level of methods.\cite{ghosh} 
Recently, the time-dependent wave packet dynamics and charge transfer processes have been reported between the one-dimensional ground state (GS)
and the low-lying excited states (ES) potential energy curves in collinear and perpendicular approaches of H towards CS$^+$ at the MRCI/aug-cc-pVQZ 
level of theory.\cite{molphys} \\

A kinetic study is important to understand the mechanisms and nature of molecule formation during collisions. 
Study of collisional rates of CS with H$^+$ ion is
important and still lacking. 
Interpretation of observed spectroscopic data by theoretical methods require kinetic models that include the study of 
rate coefficients and in turn, cross-sections of colliding species, H$^+$ $-$ CS.
An isovalent and abundant species HCO$^+$, has been explored for low-energy rotational inelastic transitions between H$^+$ - CO
(for $j \leq 4$) \cite{dhilip} and H + CO$^+$ (for $j \leq 8$)\cite{andersson} with the computation of cross-sections and 
corresponding rate coefficient details. 
The study of HCS$^+$ with its isovalent species, HCO$^+$ and N$_2$H$^+$, which are unique targets to explore 
the chemistry of molecular ions in protostellar shocks, has been done in the past to calculate the rate coefficients of the 
species in collision with He.
Later, the abundance of HCS$^+$ has been calculated using large velocity gradient approximation in radiative transfer 
code taking the collision rate coefficients calculated by Monteiro in 1984.\cite{monteiro,podio} 
Recently, Dubernet {\it et al.}\cite{dubernet} have studied the He-HCS$^+$ system by computing their potential energy surface and inelastic rate coefficients 
for $j$ $\le$ 20 in the temperature range 5 $-$ 100 K and also 
compared the results of rate-coefficients computed previously.\cite{monteiro} 
To the best of our knowledge, there are neither experimental study nor theoretical data available on inelastic low-energy rotational 
excitation and deexcitation of CS collisions with H$^+$. \\
%

%Interstellar sulfur molecules are homologues to oxygen molecules, like HCS$^+$ and HCO$^+$. 
%Similarly, CS$^+$ is an isovalent species of CO$^+$, where later has been detected in the interstellar medium. 
%Both share presence in the atmospheres and comet tails.\cite{lust}
%CS$^+$ is an important precursor and an intermediate in the formation of HCS$^+$ and also participate 
%in the formation of CS in the interstellar medium but it has not been conclusively identified there.\cite{prasad,davies}
%Kinetics of CS$^+$ with abundant neutral molecules had been studied in laboratory where formation of HCS$^+$ 
%is suggested to be an exothermic reaction using the precursor CS$^+$ ion.\cite{barassin}
%Moreover, abundance of HCS$^+$ highly depends upon abundance of CS$^+$ and therefore essential to calculate the rate coefficients 
%to further elucidate the chemistry of sulfur bearing molecules.\cite{mcallister} \\

The analysis of observed spectra and estimation of abundance often requires collisional excitation and deexcitation rate 
coefficients. 
Quantum dynamics study of ion-neutral energy transfer process widens the understanding of kinetics of interstellar medium.\cite{bodo1}
In addition, such collisions at very low energies by cooling to ultracold regime around 1 microkelvin temperatures are being 
investigated for precise spectroscopic measurements and to study the properties of molecular gases 
near quantum degeneracy.\cite{staanum,softley,nbala} This will enable fundamental advance in quantum information processing and ultracold plasmas. 
The calculations of rotational excitation and deexcitation cross-sections and corresponding rate coefficients are performed for H$^+$ - CS system.
To investigate the collisions in the extreme quantum regime where a single partial wave dominates the entire collision, 
the study of rotational inelastic excitation up to 800 cm$^{-1}$ is extended to deexcitation transitions 
from 10$^{-7}$ cm$^{-1}$ to 5000 cm$^{-1}$ for $j$=5 rotational level of CS. \\

The present work involves the collision study of H$^+$ with CS which require accurate potential energy surface (PES). 
The GS and the low-lying ES PESs of H$^+$ - CS system has been 
generated using {\it ab initio} method and the details are discussed in Sec. \ref{sec:cdet2}. 
The multipolar expansion coefficients of potentials computed using rigid-rotor PES
which is extracted from the full PES is described in Sec. \ref{sec:multi}.
Scattering details including the
calculation of cross-sections and corresponding rate coefficients 
in the rigid-rotor approximation are provided
in Sec. \ref{sec:scatdet} 
and the results obtained using vibrationally averaged potential are provided in Sec. \ref{sec:vibavg}.
Finally the summary of the present work is provided
in Sec. \ref{sec:sum}.

\section{\label{sec:cdet2}{\it Ab Initio} Potential Energy Surfaces}

\begin{figure}[ht]
\begin{center}
\includegraphics [height=0.35\textwidth]{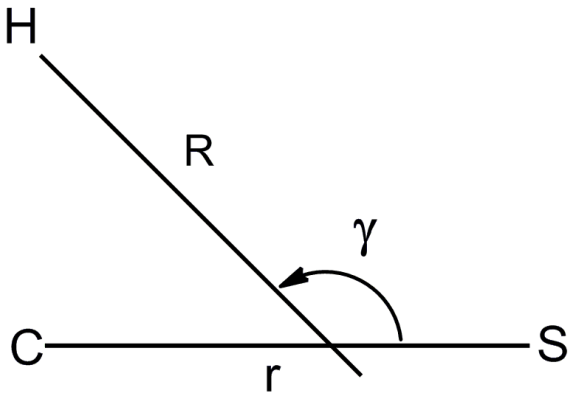}
{\caption{\label{fig:jacobi}Jacobi coordinates representation of H$^+$ collision with CS.}}
\end{center}
\end{figure}

{\it Ab initio} calculations have been performed for H$^+$ collision with CS in Jacobi coordinates to compute interaction
potential, where variables $R$ and $r$ represent the distance of
proton from center-of-mass of CS and interatomic distance of CS, respectively, and $\gamma$=$cos^{-1}$($R.r$),
is the angle between $R$ and $r$ as shown in Fig. \ref{fig:jacobi}. 
\begin{figure}[htb]
\minipage{0.50\textwidth}
  \includegraphics[width=\linewidth]{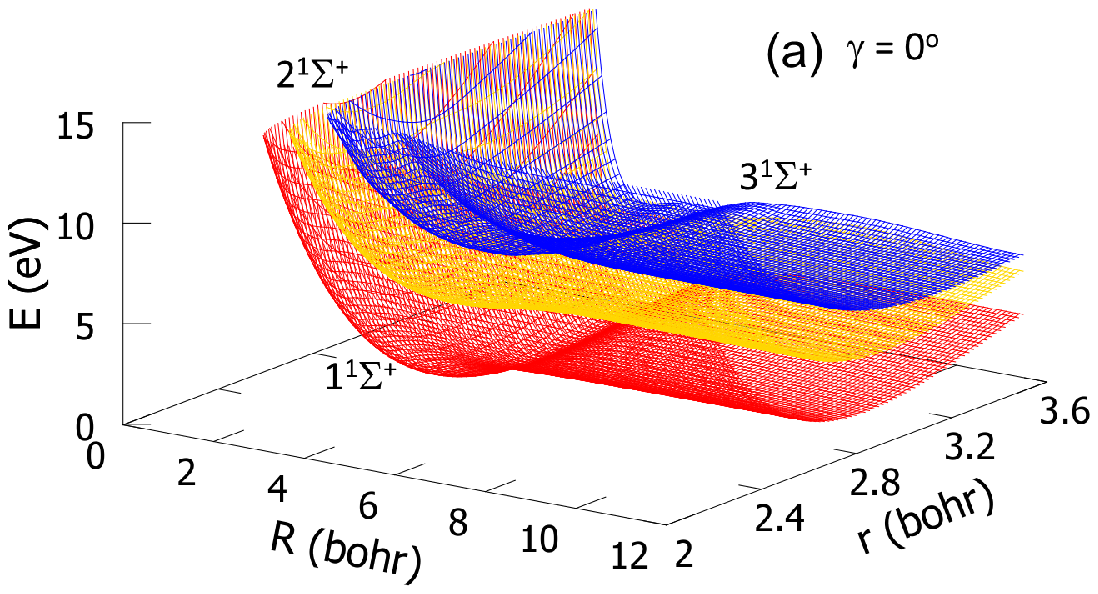}
  \includegraphics[width=\linewidth]{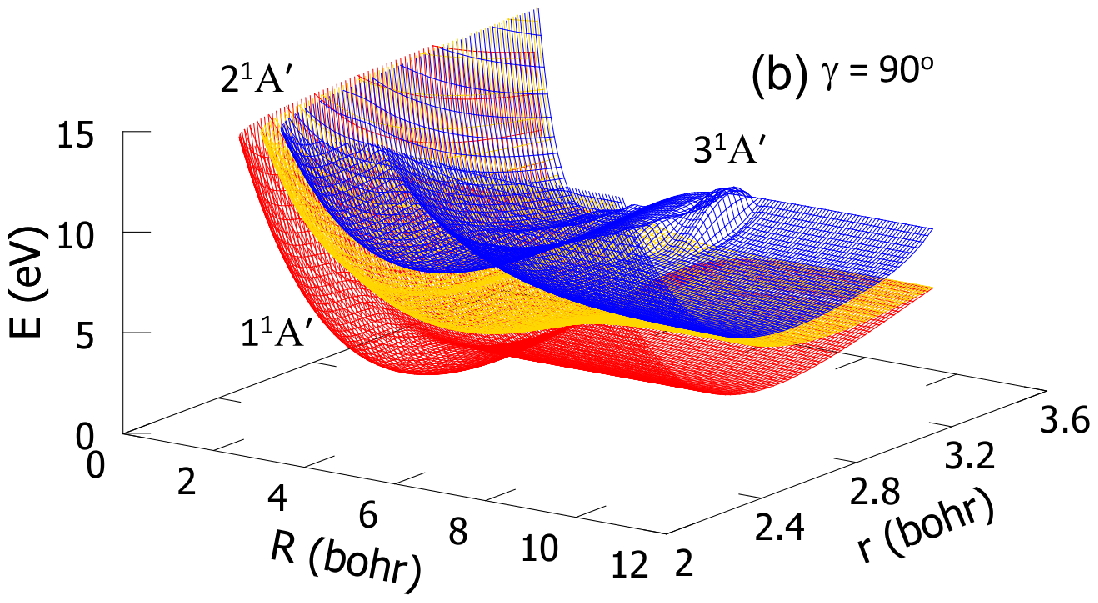}
  \includegraphics[width=\linewidth]{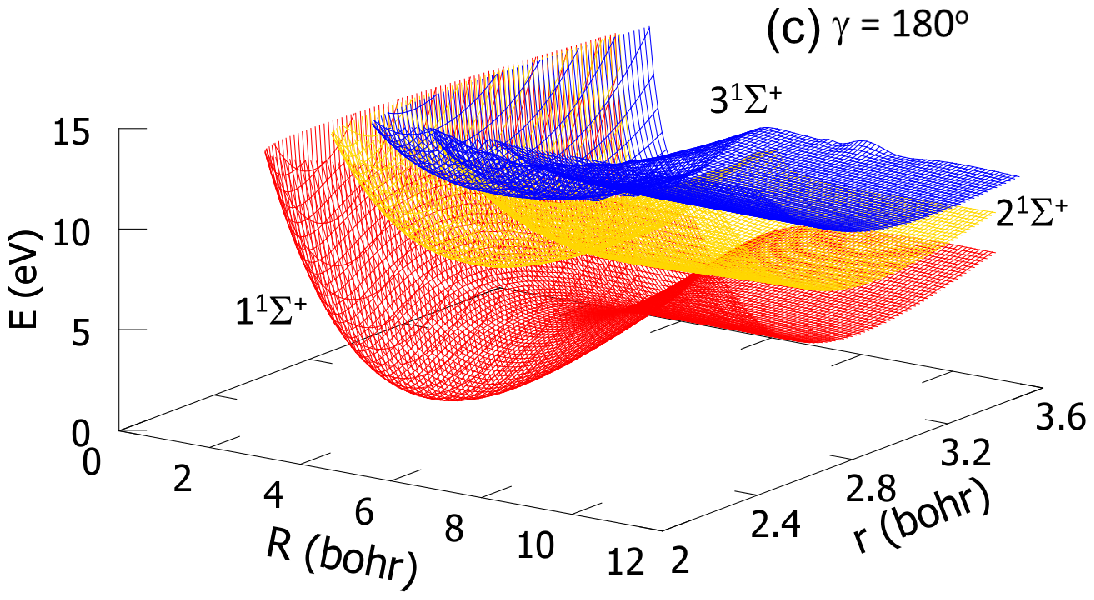}
\endminipage\hfill
\minipage{0.50\textwidth}
  \includegraphics[width=\linewidth]{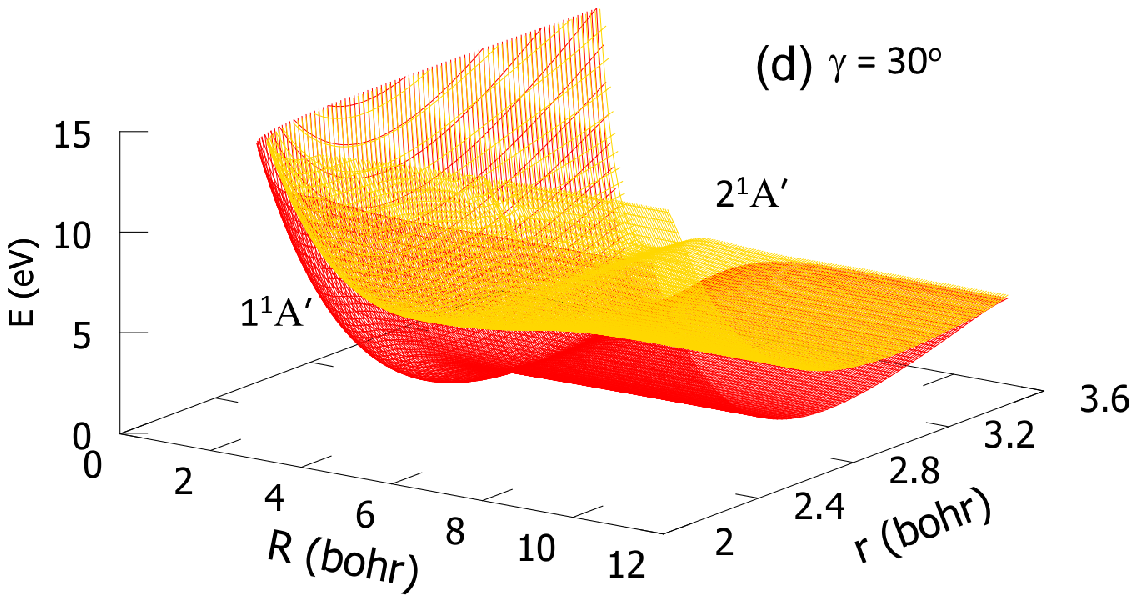}
  \includegraphics[width=\linewidth]{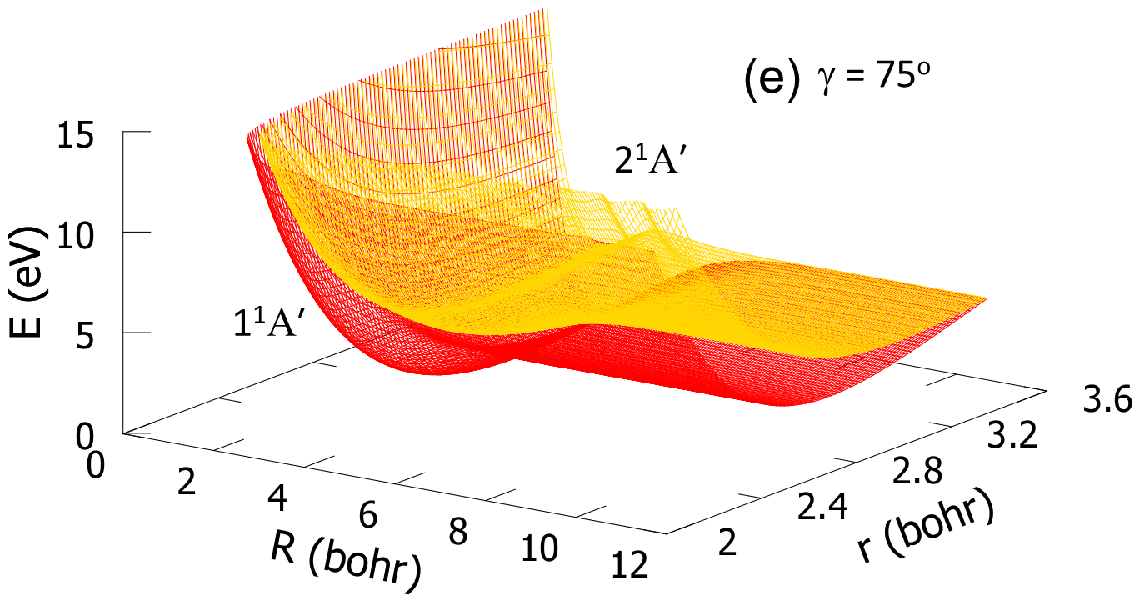}
  \includegraphics[width=\linewidth]{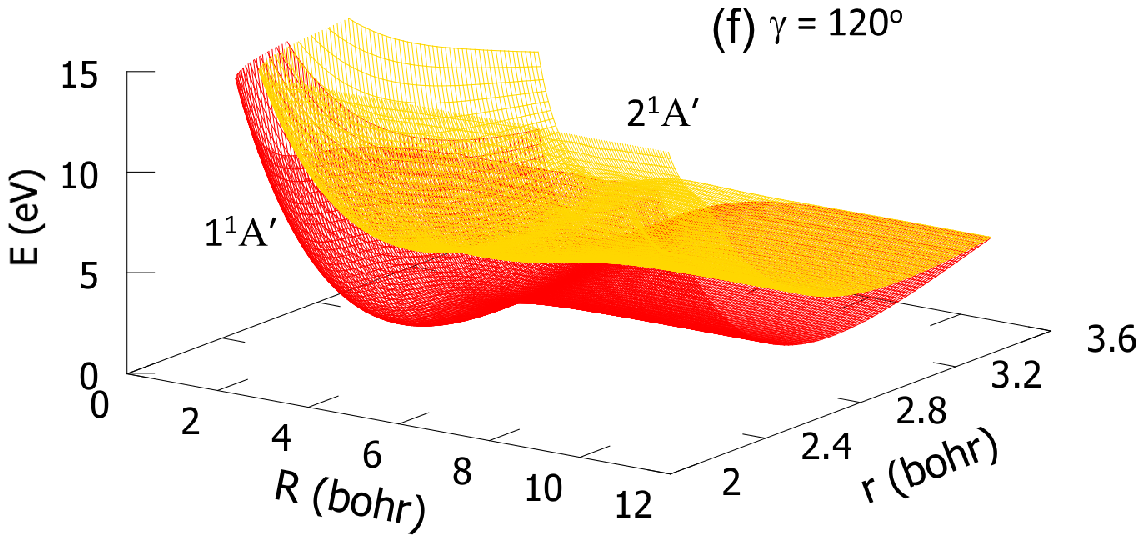}
\endminipage
\caption{\label{fig:pes}The GS (red) and the low-lying ES (yellow, blue) PESs
as a function of $R$ and $r$ at various orientations ($\gamma$).}
\end{figure}
The method employed for the PES calculation is internally contracted multireference configuration interaction (MRCI)
as implemented in MOLPRO package.\cite{molpro}
The basis sets of Dunning's augmented correlation consistent polarized quadruple zeta (aug-cc-pVQZ) for
H, C and S atoms are chosen.\cite{dunning}
The adiabatic GS and the first ES PESs are computed for orientations 0${^\circ} {\leq} {\gamma} {\leq}$ 180{$^\circ$}
with 15$^{\circ}$ increment. While the second ES PES is also computed for collinear ($\gamma$=0$^{\circ}$, 180$^\circ$) and 
perpendicular orientations ($\gamma$=90$^\circ$) as avoided crossings 
have been observed in one-dimensional PESs studied previously to investigate the non-adiabatic couplings present in the system 
using time-dependent wave packet dynamics.\cite{molphys}
For the $C_{s}$ point group, the computations employed 210 contracted functions with 133 in $a'$ and 77 in $a''$ symmetry.
The core orbitals having lowest 5 orbitals in $a'$ and one orbital in $a''$ is kept frozen and are taken from self-consistent field calculations.
The active orbitals with 6-12 $a'$ and 1-2 $a''$ symmetry incorporate remaining 10 electrons in the complete active space 
self-consistent field (CASSCF) calculations.
The MRCI calculations following CASSCF method generate reference space of 1589 configurations: N, N-1 and N-2 internal configuration
consist of 2907, 3139 and 2907 configurations, respectively. 
%This includes total number of
%contracted configurations with 2907 internal, 3139 singly external and 2907 doubly external configurations. 
Total number of contracted configurations, 2436680 include 2744 internal, 864556 singly external and 1569380 doubly 
external configurations.
The plots of computed PES as a function of $R$ and $r$ at fixed $\gamma$ for 0$^{\circ}$, 30$^{\circ}$, 75$^{\circ}$, 90$^{\circ}$, 120$^{\circ}$ and 180$^{\circ}$
orientations are shown in Figs. \ref{fig:pes}(a)-(f) as reference. 
PESs are obtained with the set of grid points as follows: $R$ = 1.4-8.0(0.2) and 8.0-11.2(0.4) $a_o$,
$r$=2.1-3.5(0.1) $a_o$ and $\gamma$ = 0$^\circ$ - 180$^\circ$(15$^\circ$).
The 1$-$3  $^1\Sigma^+$ and 1$-$3 $^1A'$ electronic states are computed for collinear ($\gamma$=0{$^\circ$}, 180{$^\circ$})
and perpendicular orientation ($\gamma$=90{$^\circ$}), respectively, while 1$-$2 $^1A'$ electronic states are computed 
for angular (off-collinear) approaches.
The numbers written in parenthesis implies the increment in the stated intervals. A total 
of 8190 {\it ab initio} points have been computed with the variation in $R$, $r$, $\gamma$ being 42, 15, 13 points, respectively. 
The obtained adiabatic surfaces have been interpolated using cubic splines and shown in Figs. \ref{fig:pes}(a)-(f). 
\subsection{Analytical Fitting of GS Surface}

The ground state potential energy surface has been fitted using the following analytical expression as a function of $R$ and $r$ 
at fixed values of $\gamma$:
%----------------------eqn-----------------------------------------------
\begin{equation}
 V(R,r;\gamma) = \sum_{i=0}^{7}\sum_{j=0}^{7-i} C_{ij}
\left (\frac{1}{R}\right)^i \left (\frac{1}{r}\right)^j 
\end{equation}\
%------------------------------------------------------------------------
The analytical fitting with the power series expansion in $R$ and $r$ has also been tried to represent zeroth order harmonic nature of CS vibration. 
Unfortunately, the function is not reproducing the potential at longer $R$ and $r$ values resulting in large error. Therefore, 
the inverse power expansion function has been tried which is reproducing the potential accurately with the  
standard deviation of the fit to be in the range of 0.5$-$4.8 meV for various $\gamma$ orientations. The fitting coefficients, $C_{ij}$, for
$\gamma$ = 0$^{\circ}$$-$180$^{\circ}$ (15$^{\circ}$) are listed in Table S1 
as supporting information.\cite{epaps} 

\subsection{Stability of HCS$^+$ and HSC$^+$ Ions}

\begin{figure}[htbp]
\includegraphics [width=4.5in]{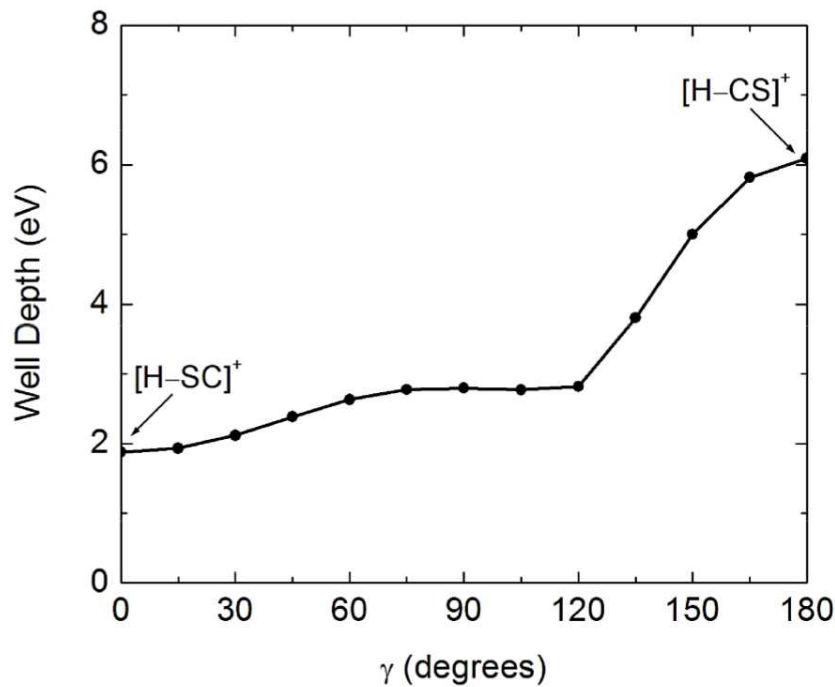}
\caption{\label{fig:welldepth}Variation in the GS potential energy well in terms of well depth as a function of
$\gamma$ from 0$^{\circ}$$-$180$^{\circ}$.}
\end{figure}
From the computed {\it ab initio} surfaces the PES profile is generated by taking the difference of energy from minimum 
of the GS potential energy well to their corresponding  
asymptotic potential for every angle ($\gamma$). A plot with calculated well depths
from HSC$^+$ ($\gamma$=0$^{\circ}$) to HCS$^+$ ($\gamma$=180$^{\circ}$), namely, from collinear through all the off-collinear 
arrangements in Fig. \ref{fig:welldepth} is shown. 
As one can observe along the minimum at each $\gamma$, there is no barrier to the rotation of H from the S-end to the C-end
of CS  and the present results are validated with results reported 
by Bruna {\it et al.} in 1978.\cite{bruna} This makes the detection of HSC$^+$ isomer in the interstellar medium unreported till date due to small well-depth of 1.88 eV. 
Moreover, the linear HCS$^+$ is highly stable relative to its isomeric linear form HSC$^+$ by 4.21 eV (97 kcal/mol) 
whereas, this difference in energy reported earlier is 110 kcal/mol. This small energy difference between the present 
and the earlier reported value is due to the choice of method 
and basis sets. Present data can be taken as more reliable as larger basis sets are employed in the calculation. 
HCS$^+$ has the minimum energy compared 
to HSC$^+$ with dissociation energy of 6.09 eV. 

\section{\label{sec:multi}Scattering Study in Rigid-Rotor Surface}

\begin{figure}[tp]
\minipage{0.60\textwidth}
  \includegraphics[width=\linewidth]{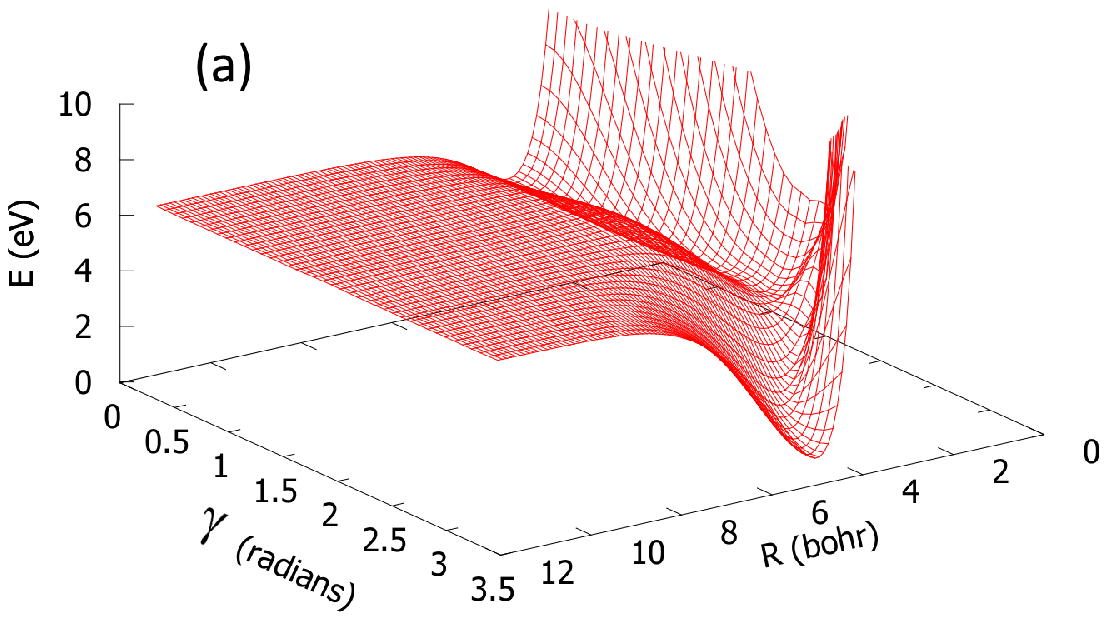}
\endminipage\hfill
\minipage{0.40\textwidth}
  \includegraphics[width=\linewidth]{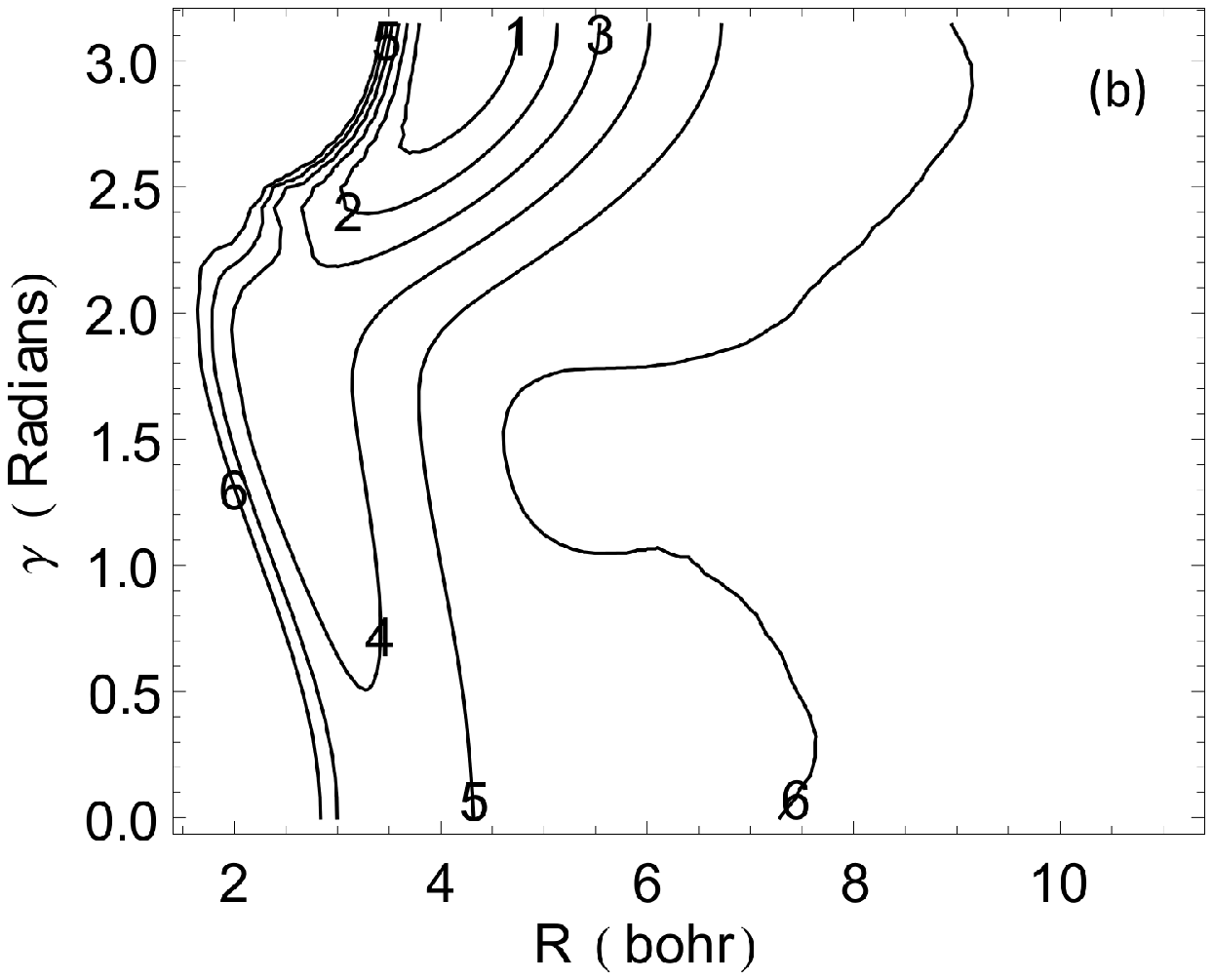}
\endminipage
\caption{\label{fig:rr}(a) Rigid-rotor PES as a function of $R$ and $\gamma$ at fixed $r_{eq}$ = 2.900 $a_{o}$ 
and (b) the contour plot of rigid-rotor surface of H$^+$ - CS system.}
\end{figure}

For the scattering study, the rigid-rotor PES has been chosen with the CS bond distance fixed at an experimental 
equilibrium distance\cite{herzberg} $r_{eq}$=2.900 $a_{o}$.
%Though the singlet GS of HCS$^+$ asymptotically corresponds to H - CS$^+$, while considering all the angular orientations 
%the rigid-rotor is fixed at the equilibrium distance of CS only. 
%
In the present work the closed channel, H$^+$ - CS ($^1\Sigma^+$), has been considered for scattering study although open channel, 
H - CS$^+$ ($^2\Sigma^+$),
is also energetically accessible. 
The rigid-rotor PES is described as a function of $R$ and $\gamma$ at $r_{eq}$ shown in Fig. \ref{fig:rr}(a). Two-dimensional contour plot 
of the surface %($1 ^{1}\Sigma^+$) 
%that asymptotically correlates to the CS$^+$($1 ^{2}\Sigma^+$) + H($^2S$) 
is shown in Fig. \ref{fig:rr}(b).
The numbers in the plot indicate the energy level of the contours in eV.
From the plots, it can be seen that the global minimum, HCS$^+$ lies at $R$ = 4.0 $a_o$ and $\gamma=180^{\circ}$ (3.14 radians). 

\subsection{Asymptotic Potentials}
Since the collision system is ionic, the asymptotic long-range potential will involve multipole moments which can be 
obtained at $r=r_{eq}$ as described below.
%The asysmptotic long-range potential of multipolar interaction at $r=r_{eq}$ is given by
%
\begin{eqnarray}
V_{as}(R,r_{eq},\gamma) \approx \frac{\mu}{R^2}P_1(cos\gamma)+\frac{Q}{R^3}P_2(cos\gamma)- \frac{\alpha_{0}}{2R^4}-\frac{\alpha_{2}}{2R^4}
P_2(cos\gamma) + O(P_3)
\end{eqnarray}
where $V_{as}$ is the asymptotic potential, $\mu, Q,$ are dipole, and quadrupole moments, respectively, and ${\alpha_{0}}$ and ${\alpha_{2}}$ are
dipole polarizability components at $r_{eq}$. $P's$ denote the Legendre polynomials.
The parameters, multipole moments and polarizability components at $r_{eq}$=2.900 $a_{o}$ are computed at the SCF and the MRCI level using MOLPRO package adopting Dunning's basis
set of aug-cc-pVQZ. 
The computed values for $\mu$ = 0.785 a.u., $Q$ = -1.820 a.u., $\alpha_{0}$ = 36.996 a.u. and $\alpha_{2}$ = 23.878 a.u. are used to obtain $V_{as}$. 
$R$ is varied from 4.0$-$200.0 $a_{o}$ for computing $V_{as}$ (long-range interaction potential) and is merged with the 
{\it ab initio} PES (short-range interaction) available in the range of $R$=1.4$-$11.2 $a_o$ using cubic-spline interpolation method. This surface has been
used for computing multipolar expansion coefficients.

\subsection{Multi-Polar Expansion Coefficients}
%
%\begin{figure} [!htbp]
%\includegraphics [width=4.6in]{V0_5.ps}
%\includegraphics [width=4.6in]{V6_12.ps}
%\caption{\label{fig:vlambda}Radial multipolar expansion coefficients for
%H$^+$ + CS system as a function of $R$ at $r$ = $r_{eq}$ with (a) the lower coefficients $V_\lambda$ = 0 to 5
%and (b) the higher coefficients $\lambda$ = 6 to 12. A few lower even coefficients show attractive wells while odd coefficients show 
%repulsive behavior with barrier.}
%\end{figure}
%
\begin{figure}[htb]
\minipage{0.65\textwidth}
  \includegraphics[width=\linewidth]{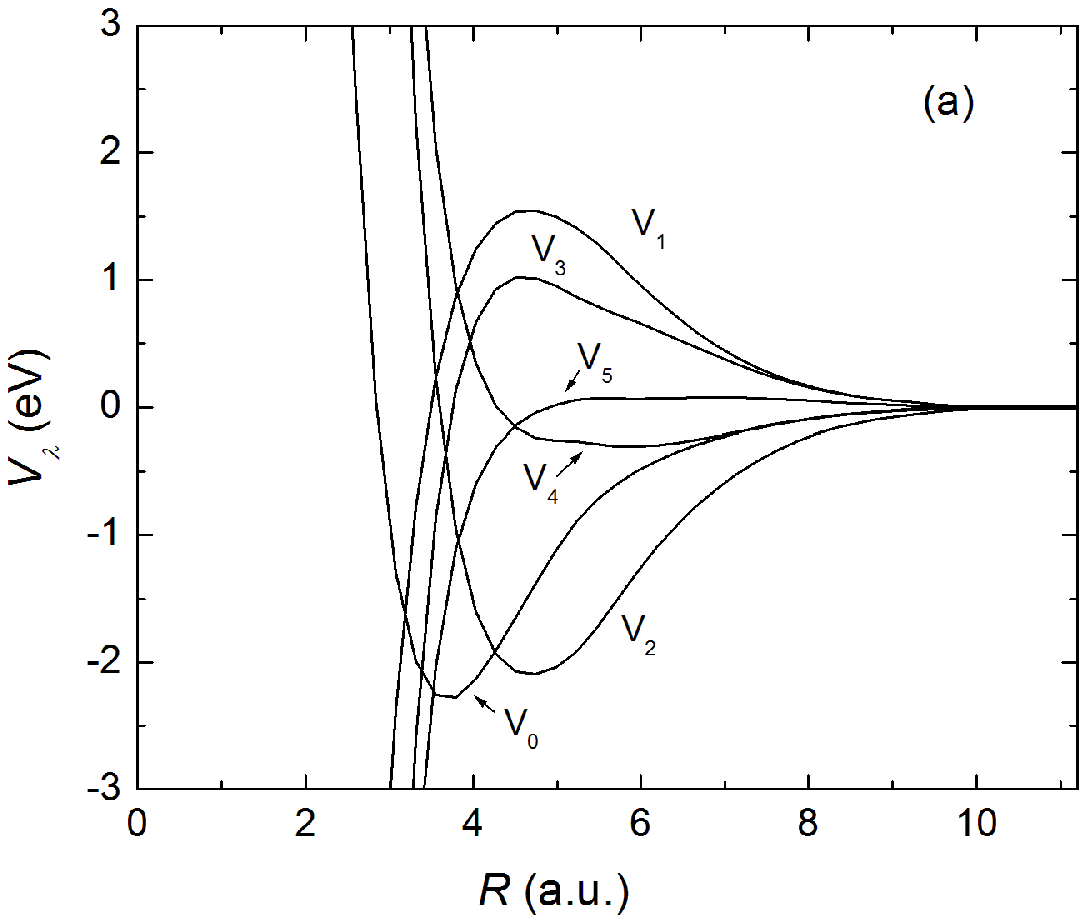}
\endminipage\hfill
\minipage{0.65\textwidth}
  \includegraphics[width=\linewidth]{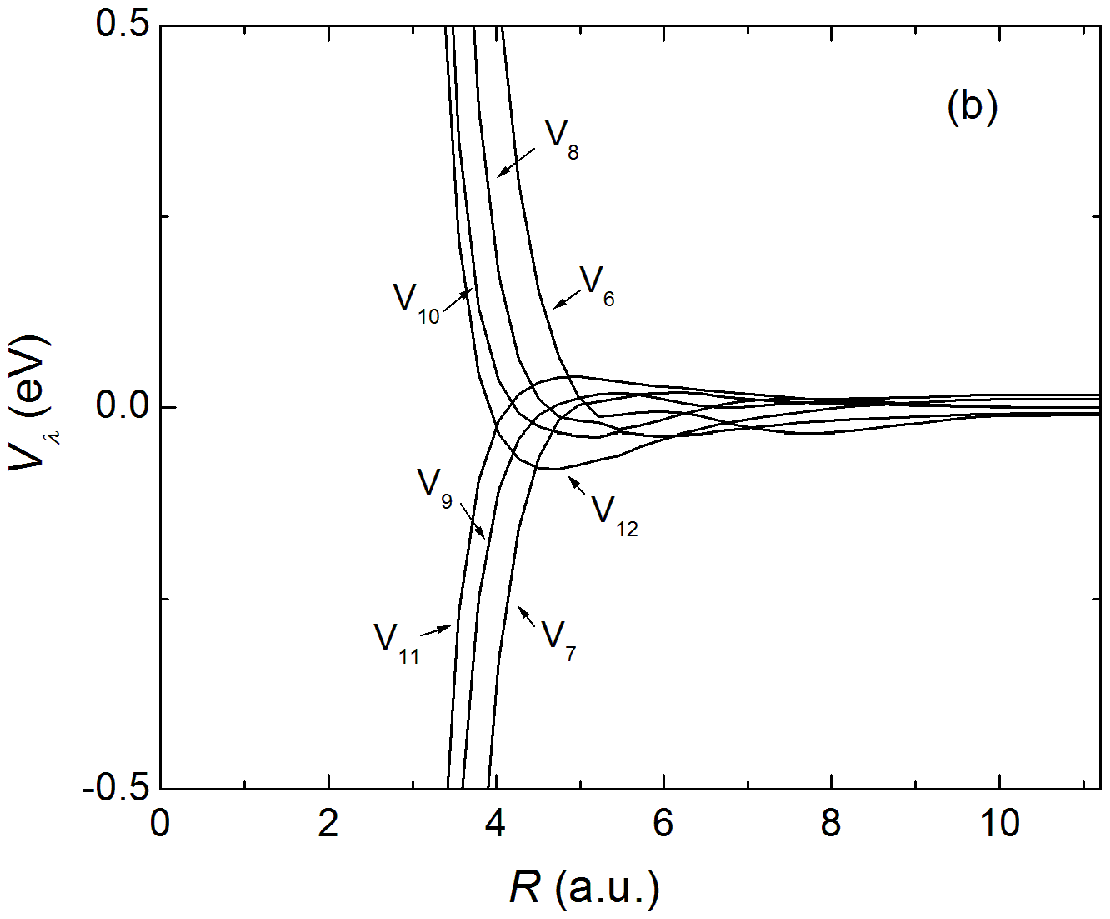}
\endminipage
\caption{\label{fig:vlambda}Radial multipolar expansion coefficients for
H$^+$ - CS system as a function of $R$ at $r$ = $r_{eq}$ with (a) the lower coefficients $V_\lambda$ = 0 to 5
and (b) the higher coefficients $\lambda$ = 6 to 12. A few lower even coefficients show attractive wells while odd coefficients show 
repulsive behavior with barrier.}
\end{figure}

The rigid-rotor surface is fitted in the expansion of Legendre polynomials for scattering calculation as follows,
\begin{equation}{\label{eqn:mp}}
V(R,r=r_{eq},\gamma) = \sum_\lambda V_{\lambda}(R)P_{\lambda}(cos \gamma)
\end{equation}
where, $P_{\lambda}'$ s are the Legendre polynomial functions.
The multipolar expansion coefficients are computed for $\lambda$ = 0 to 12 and the plots of 13 computed values of $V_{\lambda}$ 
as a function of $R$ are shown in Fig. \ref{fig:vlambda}.
It can be seen from Fig. \ref{fig:vlambda}(a) for $\lambda$=0-5 that $V_{0}$ and $V_{2}$ exhibit deep potential attractive wells. 
$V_{1}$ and $V_{3}$ display repulsive behavior with barriers. While $V_{4}$ has shallow well, $V_{5}$ show barrier-less repulsive behavior. 
The anisotropy of the interaction potential observed for $V_{\lambda}$, where $\lambda$ = 6-12 is shown in Fig. \ref{fig:vlambda}(b). 
The magnitude of the coefficients decreases as the $\lambda$ value increases. 
% for $V_{\lambda}$, where $\lambda$ = 6-12.
The coefficients obtained are interpolated using cubic spline method in the range of $R$ = 1.4$-$200 $a_0$ for scattering study. 
The computed multipolar expansion coefficients indicate anisotropic nature of the rigid-rotor surface.
The resulting $V_{\lambda}$'s are used for the kinetic study to compute various dynamical parameters such as integral and differential 
cross-sections, and rate-coefficients at low and ultracold collision energies.  % as described below.
Tabulated values of $V_{\lambda}$ fitting coefficients as a function of $R$ are provided in Table S2 ($\lambda$=0$-$5) and 
Table S3 ($\lambda$=6$-$12) as supporting information.\cite{epaps}

\section{\label{sec:scatdet}Scattering Phenomena: Close-Coupling Calculations}

The effects of anisotropy in potentials can be well studied through rotational transition cross-sections. 
To obtain the cross-sections, time-independent coupled scattering equations are solved as implemented 
in the MOLSCAT code.\cite{molscat,mcbane}
Close-coupling method\cite{dalgarno1,gianturco} (CC) involve the solution of time-independent Schr\"odinger equation which compute the cross-sections as
\begin{equation}
\sigma_{j\rightarrow j'}(E_j) = \frac{\pi}{k_{j}^2(2j+1)} \sum_{J=0}^{} \sum_{l=|J-j|}^{J+j} \sum_{l'=|J-j|}^{J+j'}
(2J+1) |\delta_{jj'} \delta_{ll'} - S_{jj'll'}^{J}(E_j)|^2
\end{equation}
where total angular momentum $\mathbf{J}$ = $\mathbf{l}$ + $\mathbf{j}$ includes orbital angular momentum of complex and 
rotational angular momentum of the diatomic molecule. 
$k_{j}$ = $\sqrt{2\mu E_{j}}/{\hbar}$ represents the wave vector for the incoming channel where $E_j$ is center of mass kinetic energy and $S_{jj'll'}$ is the scattering $S$-matrix. \\ 

%%%%%%%%%%%%%%%%%%%%%%
The inelastic differential cross-sections can be computed using,
\begin{equation}\label{eqn:dcs}
\frac{d\sigma_{j \rightarrow {j'}}}{d\omega} = \frac{1}{(2j+1)k_j^2} \sum_{m,m'}\left|\sum_{l'}(2l'+1)\left[\frac{(l'-|m-m'|)!}{(l'+|m-m'|)!}\right]^{1/2}
  A^{l'}_{(jm \rightarrow j'm')} P_{l'}^{|m-m'|}(\cos\theta)\right|^2
\end{equation}
in the center-of-mass of the system in which the initial and final
wave vectors, ${k_j}$ and ${k_{j'}}$, are related via
$\cos\theta$ = ${k_j.k_{j'}}$. The partial amplitude $A^{l'}_{(jm \rightarrow j'm')}$ is related to transition matrix, $T_{jl \rightarrow j'l'}^J$ 
which in turn is related to scattering $S$-matrix. \\ %, $S_{jl \rightarrow j'l'}^J$ \\

%%%%%%%%%%%%%%%%%%%

The rate coefficients are calculated by averaging the obtained cross-sections over a Boltzmann distribution of kinetic energy
\begin{equation}\label{eqn:rate}
k_{j{\rightarrow}j'}(T)=\sqrt{\frac{8k_BT}{\pi\mu}}\left({\frac{1}{k_BT}}\right)^2\int_0^\infty\sigma(E_j)E_j\exp{\left(\frac{-E_j}{k_BT}\right)}dE_j
\end{equation}
where $k_B$ is the Boltzmann constant and $\mu$ is the reduced mass of the system. % and $E$ is the collision energy. \\ %= E_{collision}$.\\

%The deexcitation rate coefficients can be calculated from the rate coefficients of the reverse transitions based 
%on microscopic reversibility:
%
%\begin{equation}\label{eqn:deexcit} 
%k_{j'{\rightarrow}j}(T)=(2j + 1)/(2j' + 1) \exp{\left((k_BT)^{-1}(E_{j'}-E_j)\right)} k_{j\rightarrow j'}(T)
%\end{equation}
%
\subsection{State-to-State Excitation Cross-sections and Rate Coefficients} %Low Energy Collisions}
\begin{figure} %[!ht]
\begin{center}
\includegraphics [height=0.55\textwidth]{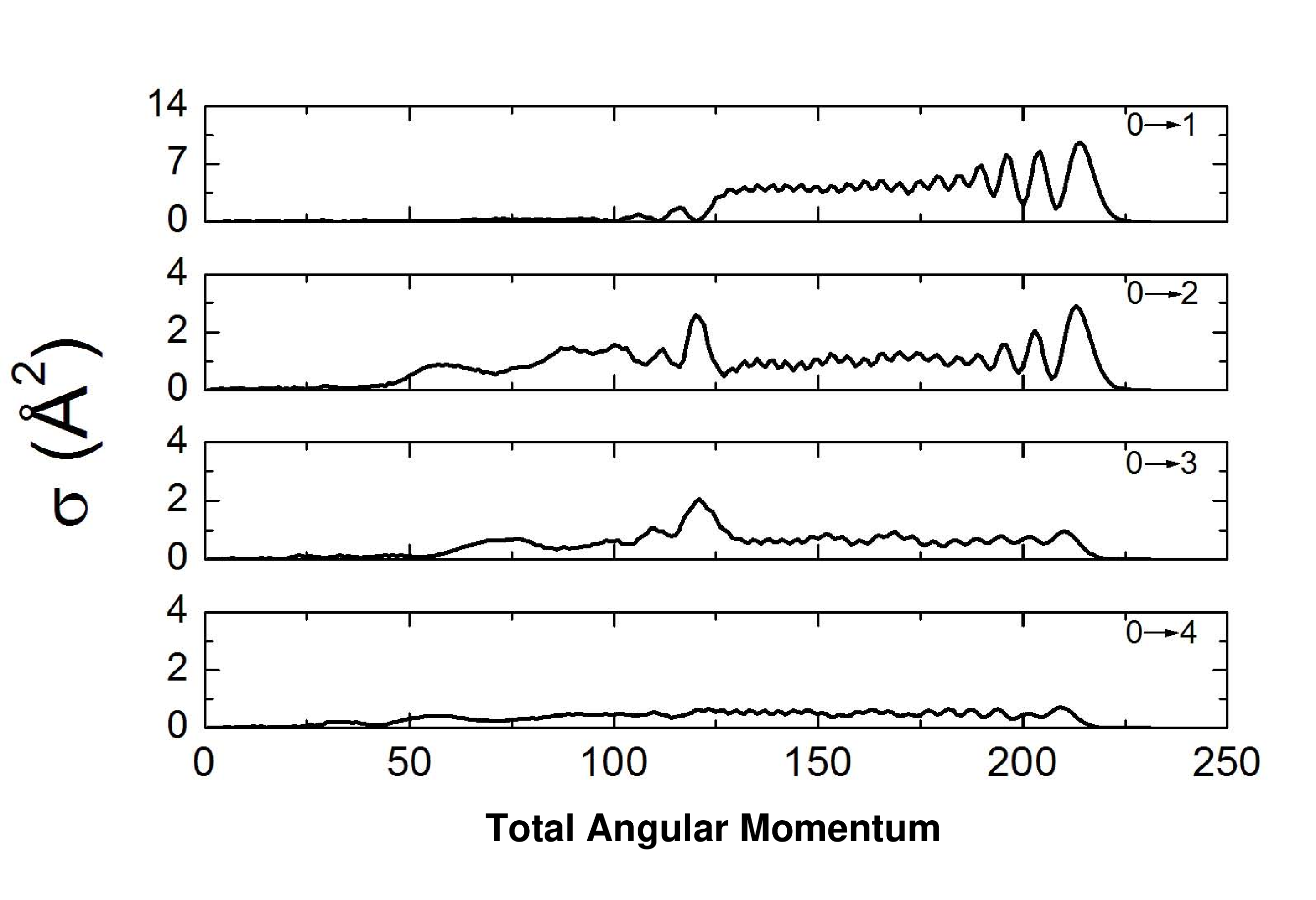}
{\caption{\label{fig:opacity}Computed partial opacities for state-to-state inelastic cross-sections as a
function of total angular momentum for the CC calculation at $E$ = 300 cm$^{-1}$.}}
\end{center}
\end{figure}
\begin{table}
\caption{\label{tab:table1}Parameters included in the MOLSCAT calculations for excitations.}
\begin{center}
\begin{tabular}{ccccccc}
\hline
Energy      & & No. of         & & Total No.     &   &  Total Angular \\
(cm$^{-1}$) & & Open Channels  & & of Channels   &   &  Momentum,  $J$\\
\hline
20 & &  5 & & 20 & & 80\\
50 & & 8 & & 23 & & 100\\
80 & & 10 & & 25 & & 120 \\
100 & & 11 & & 26 & & 150\\
200 & & 16 & & 29 & & 200\\
300 & & 19 & & 32 & & 280\\
%400 & & 22 & 30  & & \\
800 & & 31 & & 35 & & 700\\
\hline
\end{tabular}
\end{center}
\end{table}
Reliable close-coupling method is used 
to compute the cross-sections in the energy range of 2$-$800 cm$^{-1}$ employing %the log derivative method of Manolopoulos. % as implemented in the MOLSCAT.
the diabatic modified log-derivative method of Manolopoulos for radial integration of coupled channel equations.\cite{manolopoulos}
The input parameters in the calculation are taken as, CS rotational constant, $B_e$ = 0.81923 cm$^{-1}$ and 
reduced mass of the system, $\mu$, being 0.985 a.u. with values of $R_{min}$ and $R_{max}$ as 1.4 and 100 $a_0$, respectively.
The CC calculations have been performed 
from an energy value corresponding to the opening of the lowest inelastic
channel to a total energy of 800 cm$^{-1}$. The energy range has been 
carefully spanned to observe the presence of resonances due to 
quasi-bound states supported by the attractive part of the 
interaction potential. The energy steps are 0.1 cm$^{-1}$ 
below 40 cm$^{-1}$,
1 cm$^{-1}$ from 40 to 100 cm$^{-1}$, 20 cm$^{-1}$ from 100 to 400 cm$^{-1}$,
and 100 cm$^{-1}$ from 400 to 800 cm$^{-1}$. 
Minimum 13 closed channels are included at each collision energy up to 300 cm$^{-1}$ to ensure 
convergence of cross-section. Maximum value of rotational quantum number 
taken is 35 at $E$ = 800 cm$^{-1}$. %with 10 as integration step-size.
Steps parameter is kept at 30 up to 100 cm$^{-1}$ and at 10 for energies above 100 cm$^{-1}$. 
Cross-sections as a function of number of closed channels for $j$=0 to $j'$=1 excitation at different energies is shown in 
Fig. S1 of supporting information.\cite{epaps}
Also, convergence of cross-sections are achieved through sufficient number of partial waves or total angular momentum. 
To test the convergence, % of cross-section values, 
inelastic opacities have been computed for the lowest four channels as a function of 
total angular momentum and shown for the collision energy $E$=300 cm$^{-1}$ in Fig. \ref{fig:opacity}.
The magnitude of opacity is quite large for ($0 \rightarrow 1$) transition and
it decreases in the order from ($0 \rightarrow 2$), ($0 \rightarrow 3$) and ($0 \rightarrow 4$). 
The convergence is ensured at values of
angular momentum (maximum) of 280 when the collision energy is 300 cm$^{-1}$. 
The parameters used in the calculation are provided in the Table \ref{tab:table1}. 
The calculations are performed using parallel code\cite{mcbane} of MOLSCAT for energies above 400 cm$^{-1}$. \\

\begin{figure}
\begin{center}
\includegraphics [width=4.8in]{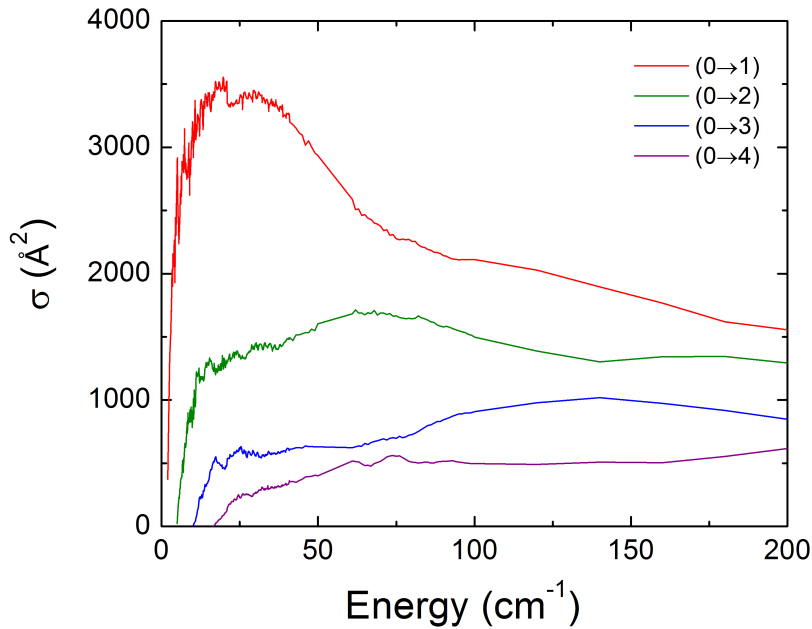}
%\includegraphics [width=4.8in]{cc_800.ps}
%{\caption{\label{fig:sigmacc}Computed state-to-state integral cross-section using the CC method (a) up to $E$ = 200 
%cm$^{-1}$ and (b) up to $E$ = 800 cm$^{-1}$.}}
{\caption{\label{fig:sigmacc}Computed state-to-state integral cross-section for excitations from $j$=0$\rightarrow$$j'$=1, 2, 3, 4 
using the CC method up to $E$ = 200 cm$^{-1}$.}}
\end{center}
\end{figure}
%
%
%\begin{figure}[!ht]
%\begin{center}
%\includegraphics [height=0.80\textwidth]{cc_cs_sigma_j.ps}
%{\caption{\label{fig:sigmaj}Computed state-to-state integral cross-section as a function of $j'$ states at $E$ =
%(a) 50 cm$^{-1}$, (b) 80 cm$^{-1}$, (c) 160 cm$^{-1}$ and (d) 200 cm$^{-1}$ for the CC and the CS calculations.}}
%\end{center}
%\end{figure}
%
Full close-coupling calculations have been performed at low energies  (2$-$800 cm$^{-1}$) and  %as explained above. 
Fig. \ref{fig:sigmacc} displays the variation of calculated 
rotationally inelastic excitation cross-sections ($\sigma_{j\rightarrow j'}$) as a function of collision energy up to 200 cm$^{-1}$. 
The rotational excitations computed are plotted for 
$j$=0 $\rightarrow$ $j'$ = 1$-$4. Close examination of the cross-sections at low energies typically below 50 cm$^{-1}$ show resonances. 
The cross-sections oscillate at low energies attaining maximum while with the increase of energy the magnitude of cross-sections decreases and 
plateau is reached.
Maximum value of cross-section is found in the transition from $j$ = 0$ \rightarrow$ $j'$ = 1, 
which decreases for higher rotational energy levels %from higher to lower 
in the order of $j$ = 0 $\rightarrow$ $j'$ = 2$-$4. For all the excitations, cross-sections follow the similar behavior.  
The calculations involving close coupling method for excitations are performed with energy restricted up to 800 cm$^{-1}$ 
as convergence of cross-sections put limitation with the increase of open channels and value of total angular momentum.
This is due to collision energy increase, the number of $j$ levels coupled by the potential increases and the number of close-coupling equations 
to be solved become very large. \\

\begin{figure}[tp]
\begin{center}
\includegraphics [height=0.65\textwidth]{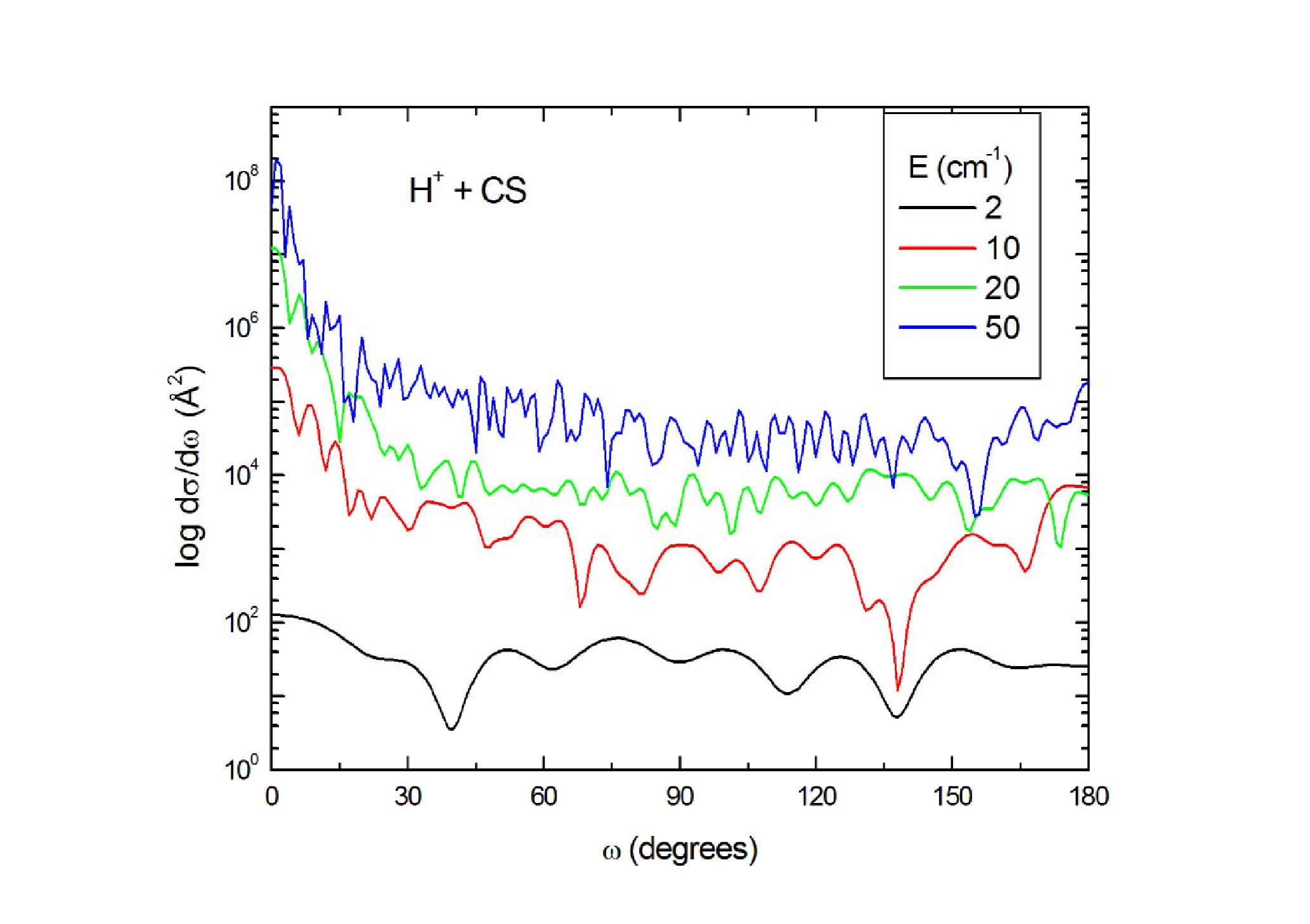}
{\caption{\label{fig:dcsfig}Differential cross-sections for $j$ = 0 $\rightarrow$ $j'$ = 1 excitation at the range 
of collision energies below 50 cm$^{-1}$.}}
\end{center}
\end{figure}
\begin{figure}[tp]
\begin{center}
\includegraphics [height=0.65\textwidth]{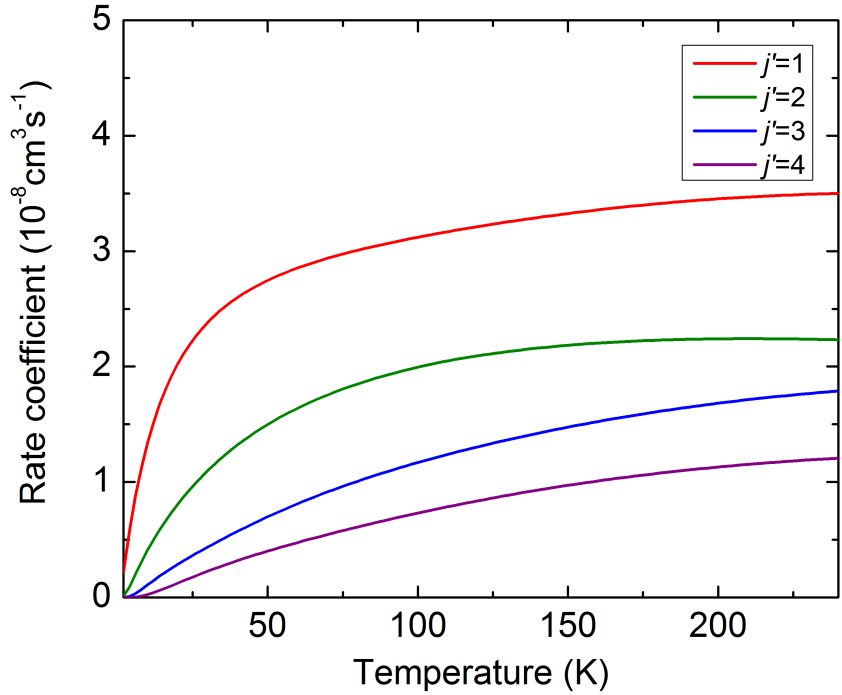}
{\caption{\label{fig:sigmarate}Computed state-to-state rate-coefficient for excitation from $j$=0$\rightarrow$$j'$=1, 2, 3, 4 using the CC method.}}
\end{center}
\end{figure}
%
%\subsection{Differential Cross-Sections}
The CC $S$-matrices obtained from MOLSCAT are used to 
compute state-to-state differential cross-sections, 
d$\sigma$/d$\omega$, 
at intervals of 1$^\circ$ using the Eq. (\ref{eqn:dcs}). 
The results are shown in Fig. \ref{fig:dcsfig} which shows
the angular scattering for the transitions
$j$ = 0 to $j'$= 1 at different energy values, namely, $E$ = 2, 10, 20 and 50 cm$^{-1}$.
For clarity, the differential cross-sections at increasing energies have been shifted by one order of magnitude 
higher correspondingly. It is evident from the figure that the oscillations in the differential cross-section is less
at lower energy while it gradually increases attaining maximum at 50 cm$^{-1}$. \\

The values of the cross-section allow one to calculate rate-coefficients as function of temperature using the Eq. (\ref{eqn:rate}).
The state-to-state rate coefficients are computed for rotational transitions 
over a range of temperatures up to 240 K as shown in Fig. \ref{fig:sigmarate}. 
The rates are higher for (0 $\rightarrow$ 1) transitions similar to cross 
sections and decreases for other higher excitations.
In the spanned range of temperatures, rates are
found to be large in magnitude for (0 $\rightarrow$ 1). (0 $\rightarrow$ 2) 
excitation rate is found to be two-third in magnitude of (0 $\rightarrow$ 1) while 
(0 $\rightarrow$ 2) rate is one-half of (0 $\rightarrow$ 1) magnitude. 
%Rates attain maximum between 35 K to 50 K for (1$-$3)
%

%\subsection{Ultra-Cold Collisions}
\subsection{State-to-State Deexcitation Cross-Sections and Rate Coefficients}
\begin{figure}[tp]
\begin{center}
\includegraphics [height=0.65\textwidth]{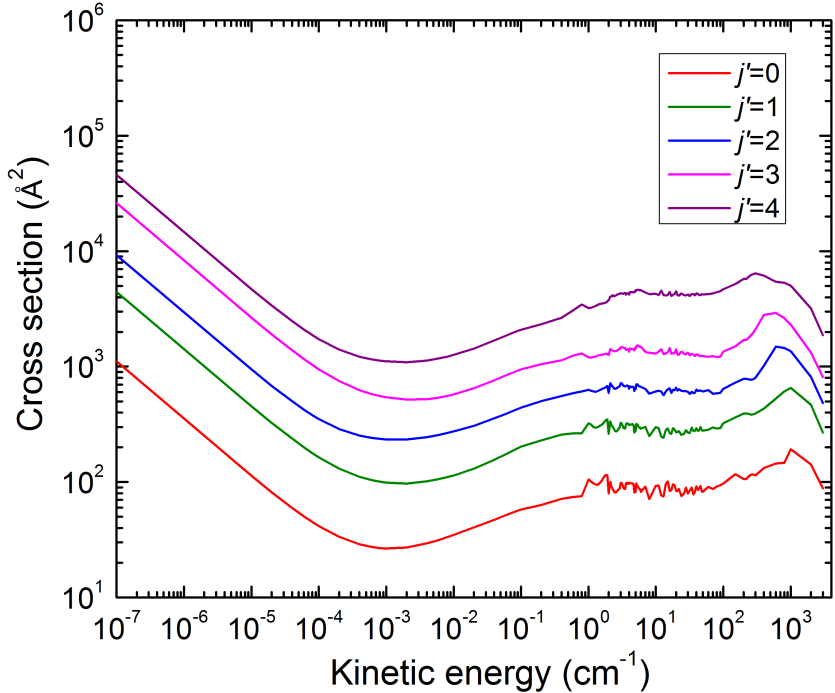}
{\caption{\label{fig:ultracold1}
Deexcitation cross-sections as a function of kinetic energy
for $j$=5$\rightarrow$$j'$=0, 1, 2, 3, 4 transitions.}}
\end{center}
\end{figure}
\begin{figure}[!ht]
\begin{center}
\includegraphics [height=0.65\textwidth]{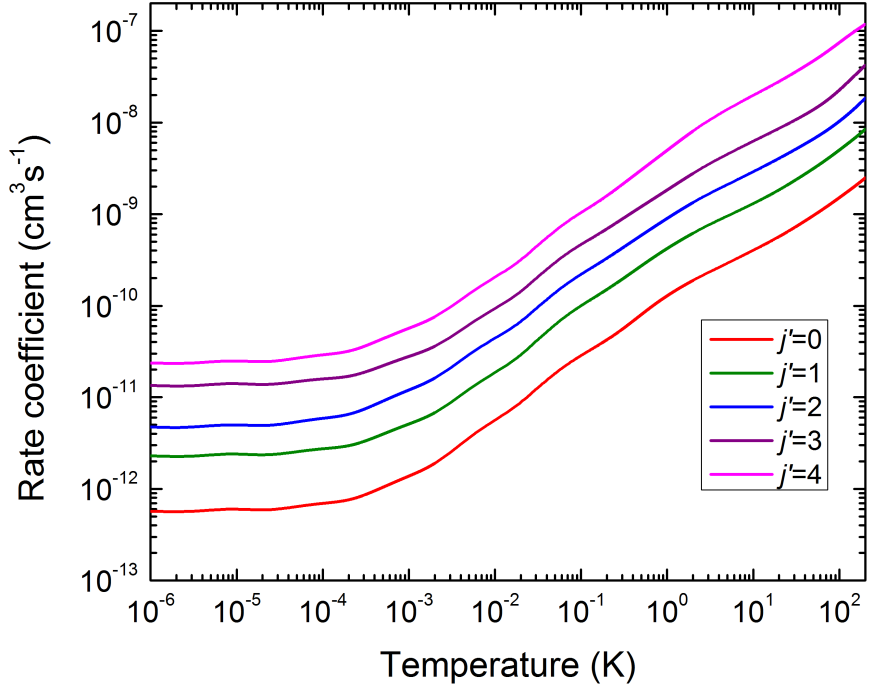}
{\caption{\label{fig:ultracold2}
Deexcitation rate coefficients as a function of temperature 
for $j$=5$\rightarrow$$j'$=0, 1, 2, 3, 4 transitions.}}
\end{center}
\end{figure}
State-to-state cross-sections have been obtained for rotational deexcitation of CS molecule for initial level of 
$j$=5 to final lower rotational levels $j'$=0, 1, 2, 3, 4. 
The CC method is computationally expensive for higher energies and the coupled state approximation method 
results does not overlap with the CC result in this system. Therefore, state-to-state deexcitation cross-sections have been calculated 
up to 5000 cm$^{-1}$ using the CC method.
For the computation of cross-sections at ultra low energies from 10$^{-7}$ cm$^{-1}$ the logarithmic energy grid has been used. 
The hybrid modified log-derivative Airy propagator of Alexander and Manolopoulos\cite{alexander} which 
uses the diabatic modified log-derivative method at short range, while changes to the Airy propagator at long range is employed.
For the ultra-cold collision regime, $R_{max}$ is extended up to 200 $a_o$ with $R_{mid}$=110 $a_o$ with 30 as steps parameter in the input.
Deexcitation cross-section from kinetic energy 10$^{-7}$ cm$^{-1}$ to 5 $\times$ 10$^3$ cm$^{-1}$ for $j$ = 5 to $j'$ = 4$-$0 has been computed as a function of 
kinetic energy and shown in Fig. \ref{fig:ultracold1}. 
Parallel code\cite{mcbane} of MOLSCAT has been used to compute cross-sections for the collision energy range 
of 400 cm$^{-1}$$-$5000 cm$^{-1}$.
It is observed that $\Delta$$j$ = -1 dominates rotational quenching from higher 
rotational level to all lower possible rotational energy levels. With the increase of $j'$, 
the magnitude of cross-section increases and it is maximum for $j'$ = 4. Resonances are seen in the 
energy range from  2$-$50 cm$^{-1}$ and these gets suppressed with increasing rotational energy levels. 
In the ultra low energy regime, the cross-section is found to vary inversely to the velocity below 
10$^{-4}$ cm$^{-1}$.
This behavior is in accordance as predicted by Wigner threshold laws\cite{wigner} at ultra-low collision energies 
where only the $s$-wave scattering contributes and the cross-section vary inversely with the relative velocity. 
The deexcitation cross-sections decrease to a minimum near 10$^{-3}$ cm$^{-1}$ and then increases to reach maximum at 
10$^{-7}$ cm$^{-1}$. \\

The rotational quenching rate-coefficients for
j=5 $\rightarrow$ $j'$=4$-$0 as a function of temperature have been obtained by averaging the cross-section over Boltzmann 
distribution of kinetic energy and 
shown in Fig. \ref{fig:ultracold2}. The temperature is varied from 10$^{-6}$ K to 200 K. 
The collision rates found to be low 
and constant up to 10$^{-4}$ K then increases gradually with increasing temperature for all transitions. The deexcitation rate from 
$j$ = 5 to $j'$ = 4 is found to be maximum and decreases gradually for each $j'$ level with the lowest for $j'$=0.
This shows that $\Delta$$j$=$-$1 transition is dominant among all. The reported state-to-state rate coefficients 
for collisional excitation and deexcitation will help in describing dynamics of energy transfer process and interpretation of 
microwave observations of the interstellar gas. 
To our knowledge, unfortunately, there has been no experimental data available of rate coefficient for rotational transitions 
of CS-H$^+$ collision system. 
It is presumed that in future the availability of experimental data 
will give credence the theoretical reported rate coefficients.  \\

In the ultracold temperature region, total deexcitation rate coefficient is found to be 1.8 $\times$ 10$^{-09}$ cm$^3$ s$^{-1}$ for 
the initial $j$=5 state. 
From the total deexcitation rate coefficient the mean lifetime ($\tau$) of the rotational excited state of the CS in the H$^+$ environment 
has been estimated.
A qualitative estimate of the typical quenching lifetime expected under trap condition can be obtained from the 
simple unimolecular kinetic equation,
\begin{equation}
\frac{d}{dt} N^{(CS)^*}(t) = -\lambda N^{(CS)^*}(t)
\end{equation}
where N$^{(CS)^*}$(t) is the number density of the excited rotational states of the CS molecule present at time $t$, 
$\lambda$ = $k$(T).N$_0$$^{H^+}$ describe an effective unimolecular decay rate, 
$k$(T) is effective temperature-dependent quenching rate coefficient, N$_0$$^{H^+}$ is the H$^+$ number density.\cite{caruso}
The mean lifetime of CS present in the trap is then given by $\tau$ = 1/$\lambda$ = 1/($k$(T).N$_0$$^{H^+}$). 
Assuming a typical number density of H$^+$ = 10$^{16}$ cm$^{-3}$ and at ultracold temperature in the range of microK 
with $k$ = 1.8 $\times$ 10$^{-09}$ cm$^3$ s$^{-1}$ for initial $j$ = 5 state, the mean lifetime, $\tau$, 
of a typical rotationally excited state CS 
in the trap will be 550 ns, a time interval which corresponds 
to an effective decay rate, $\lambda$, 1.8 $\times$ 10$^7$ s$^{-1}$.
To our knowledge, there is no
rate coefficient data reported for rotational transitions taking place ranging from ultracold to low energy regions of H$^+$ 
collision with CS molecule. 
In addition, the life-time and decay rate data obtained at the ultracold region will help to 
arrive at precise spectroscopic measurements and to study the properties of molecular gases
near quantum degeneracy.

\section{\label{sec:vibavg}Scattering Study in Vibrationally Averaged Potential}
\begin{figure}[!htb]
\minipage{0.75\textwidth}
  \includegraphics[width=\linewidth]{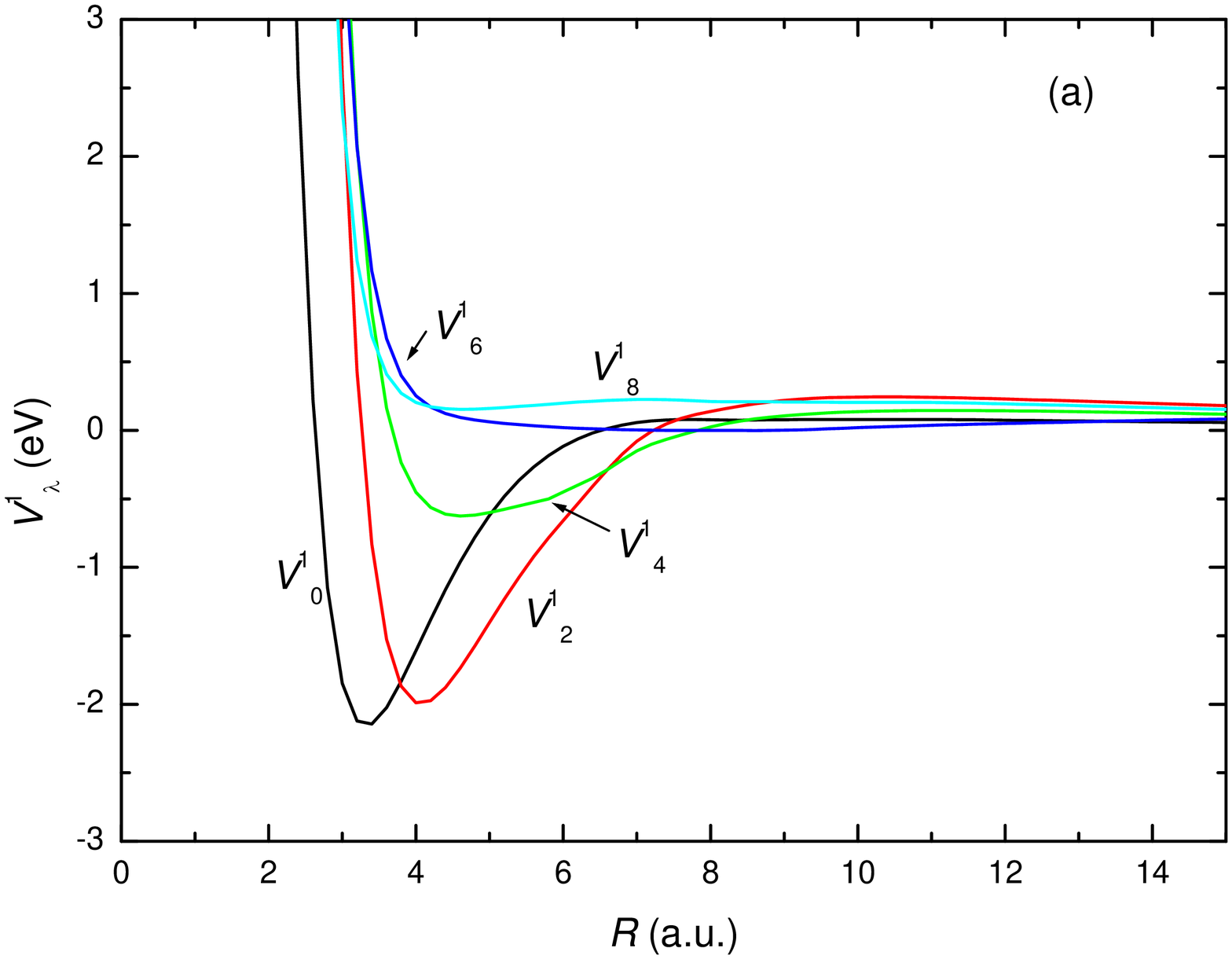}
\endminipage\hfill
\minipage{0.75\textwidth}
  \includegraphics[width=\linewidth]{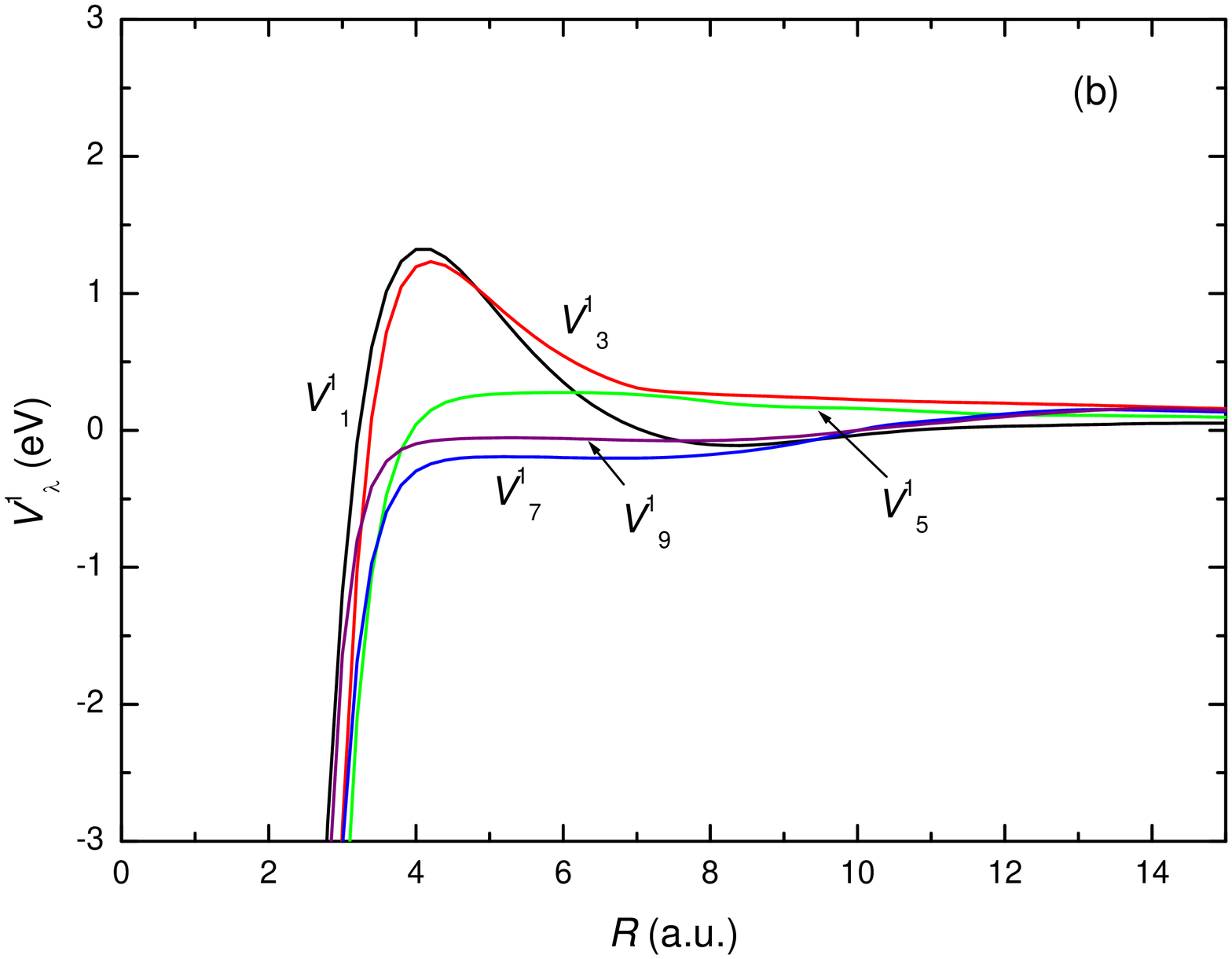}
\endminipage
\caption{\label{fig:v05_v11}Radial multipolar expansion coefficients
for vibrationally averaged potential as a function of $R$ with (a) the even 
and (b) odd coefficients.}
\end{figure}

\begin{figure}[!ht]
\begin{center}
\includegraphics [height=0.80\textwidth]{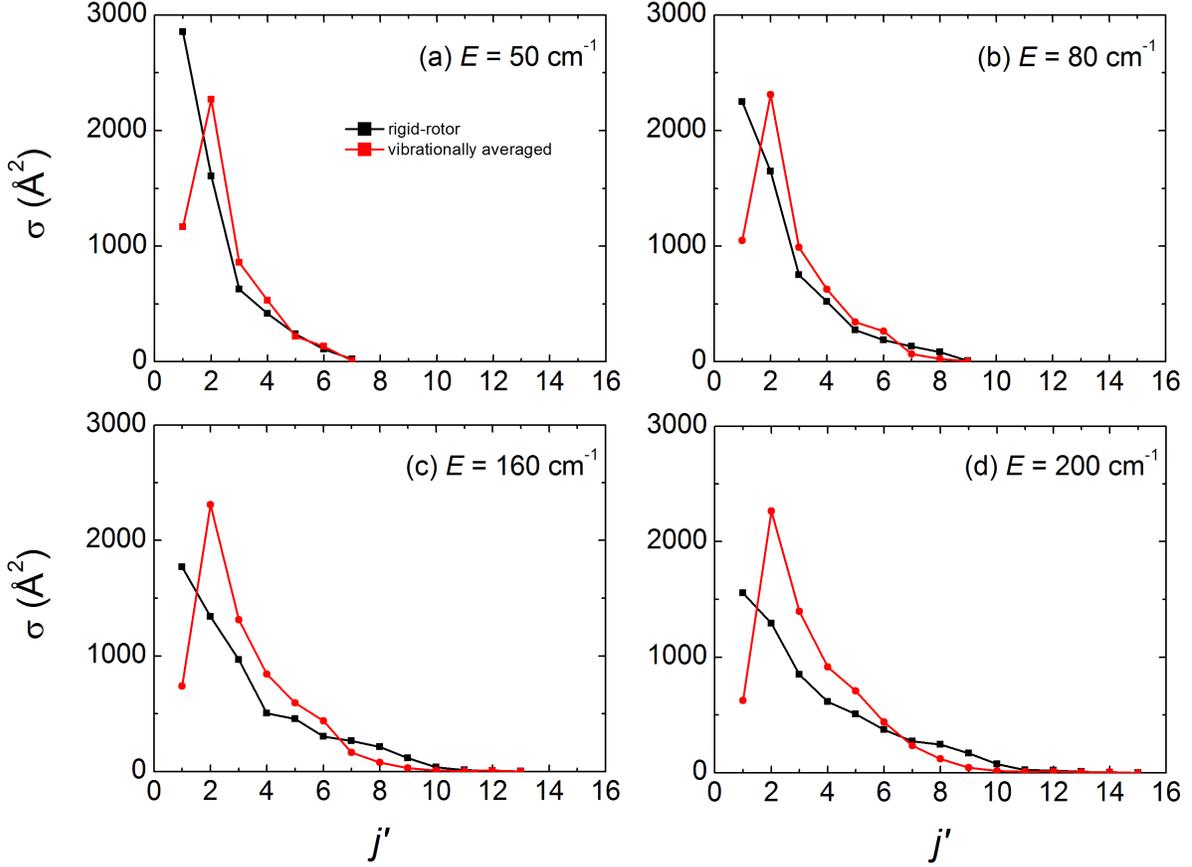}
{\caption{\label{fig:sigmajvibavg}
Comparison of the computed integral cross-sections for rigid-rotor potential and vibrationally averaged potential into all allowed values of $j'$ at $E$ = 
(a) 50 cm$^{-1}$, (b) 80 cm$^{-1}$, (c) 160 cm$^{-1}$ and (d) 200 cm$^{-1}$ }}
\end{center}
\end{figure}
The rotational excitation cross-section using 
vibrationally averaged potential %with $v$ = 1
have been calculated for the system.
The %anisotropy has been analyzed for the
vibrationally averaged potentials have been
obtained using the corresponding vibrational wavefunctions as shown below:
%The vibrationally averaged potentials are obtained as,
%
\begin{eqnarray}
V^{1}(R;\gamma) = \langle \psi_v | V(R,r;\gamma | \psi_v \rangle,
\end{eqnarray}
where $V^1$ defines the averaged potential obtained for vibrational
state $v$, with the vibrational wavefunction, $\psi_v$. 
The anisotropy of $V^1$ 
obtained is analyzed in terms of multipolar expansion coefficients, $V^1_{\lambda}$, obtained using Eq. (\ref{eqn:mp})
and shown as a function of $R$ in Fig. \ref{fig:v05_v11}.
Vibrationally averaged $V^1_{\lambda}$ terms are similar 
from those obtained by employing the rigid-rotor
potential $V_{\lambda}$ at $r$ = $r_{eq}$. 
$V_0^1$$-$$V_5^1$ are strong in magnitude and their characteristics 
are almost similar to the rigid-rotor $V_{\lambda}$'s as 
shown in the figure. 
However, $V_1^1$ and $V_3^1$ terms shows repulsive maximum at larger radial distance 
with centrifugal barrier and becomes attractive by crossing the radial axis at shorter $R$ values. 
Higher $V_{\lambda}^1$ potentials exhibit similar behaviour at larger 
radial distances as observed for the rigid-rotor $V_{\lambda}$ potentials.
Tabulated values of $V^1_{\lambda}$ fitting coefficients as a function of $R$ are provided in Table S4 ($\lambda$=0$-$5) and
Table S5 ($\lambda$=6$-$12) as supporting information.\cite{epaps}\\
%
%\begin{figure}[ht]
%\begin{center}
%\includegraphics [height=0.65\textwidth]{dcs2.ps}
%{\caption{\label{fig:dcsfig2}Differential cross-section for $j$ = 0 $\rightarrow$ $j'$ = 1, 2, 3 excitations
%at $E$ = 10 cm$^{-1}$.}}
%\end{center}
%\end{figure}
%

The integral cross-section has been computed using vibrationally
averaged potential $V_{\lambda}^1$ under CC scheme at the
collision energies, $E$=50, 80, 160 and 200 cm$^{-1}$ and shown in Fig. \ref{fig:sigmajvibavg} 
as a function of $j'$. For comparison, the cross-sections computed using rigid-rotor potential is also
shown. 
The rotational cross-sections computed using
$V^1_{\lambda}$ potential at various energies is found to be larger in magnitude as 
compared to those obtained from using the rigid-rotor potential, $V_{\lambda}$, for 
the lower $j'$ states and merging together for higher $j'$s states. 
Rotational rainbow maximum is observed at $j'$$=$2 for the cross-sections computed at various 
collision energies using vibrationally averaged potentials.\cite{schinke,bowman}

\section{\label{sec:sum}Summary and Conclusions}

{\it Ab initio} full PESs of H$^+$-CS system have been generated using the MRCI/aug-cc-pVQZ method.
The GS PES computed for various angular orientations reveals the stable configuration of
linear HCS$^+$ with dissociation energy of 6.09 eV. Potential energy profile of the system displays the barrier-less rotation
of H from the S-end to the C-end of CS.  
Multipolar expansion coefficients computed from the surface indicate anisotropic nature of the system.
Inelastic rotational excitations and deexcitations have been studied in the H$^+$ - CS system in the GS at low energy 
(2$-$800 cm$^{-1}$) and ultra-low energy (10$^{-7}$ to 5 $\times$ 10$^3$ cm$^{-1}$) regime on the rigid rotor PES 
extracted from the full PES at $r_{eq}$=2.900 $a_o$.
Cross-sections are computed for inelastic rotational transitions using 
the CC method. 
%The obtained results of rotational excitations are compared with the CS approximation. It is found that the CS 
%approximation under-predicted the cross-section compared to the CC method. Corresponding rate coefficients are calculated 
%for CC method. 
The rotational transitions favor $\Delta$$j$=+1 and $-$1 for excitation and deexcitation, respectively. 
Wigner's threshold law is obeyed for energies less than 10$^{-4}$ cm$^{-1}$ where the magnitude of cross-section increases 
as the collision energy in ultracold region is decreased. 
Rate coefficients calculated for range of energies will help in interpretation of rates of formation and decomposition of astronomical species. 
%The rotational cross sections have also been obtained using vibrationally averaged potential showed rotational rainbow maximum for $j'$=2 state.
Scattering dynamics has been performed using vibrationally averaged potential of CS molecule. The anisotropy for the 
vibrationally averaged potential has been analyzed in terms of multipolar expansion coefficients. 
The rotational cross-sections obtained using vibrationally averaged potential showed rotational rainbow maximum for $j'$=2 state.
An estimate of mean lifetime of the trapped CS due to H$^+$ collision in microkelvin region is found to
be very long time (550 ns) which will allow precise spectroscopic measurement and 
to study the properties of molecular gases near quantum degeneracy.

\section*{ACKNOWLEDGEMENTS}
This research is supported by the Department of Science and Technology, New Delhi (DST Grant No. EMR/2014/000017).
Kaur acknowledges IIT Ropar for research fellowship. The calculations are carried out in IIT Ropar High-Performance Computing cluster facility. 
%Thanks to Prof. N. Balakrishnan for fruitful discussions on ultracold collisions.

%\section*{Supporting Information Available}
%The cross-sections as a function of number of closed channels for
%$\gamma$ = 0$^{\circ}$$-$180$^{\circ}$ (15$^{\circ}$) are shown in Fig. S1. 
%The tabulated $C_{ij}$ fitting coefficients in Table S1, $V_{\lambda}$ coefficients in Tables S2-S3 and 
%$V^1_{\lambda}$ coefficients in Tables S4-S5
%are provided as supporting information.
%This information is available free of charge via the Internet at http://pubs.acs.org/.
  
\newpage
%\bibliography{refs}
\bibliographystyle{jasasty.bst}

\end{document}